\documentclass[natbib]{svmult}

\usepackage{amsmath,amssymb,graphicx,makeidx,multicol,url}
\usepackage[bottom]{footmisc}

\newcommand{\bvec}[1]{\ensuremath{\mbox{\boldmath $\mathbf{#1}$}}}

\makeindex

\begin{document}


\title*{What can Lattice QCD theorists learn from NMR spectroscopists?}

\author{George T.\ Fleming\inst{1}}

\institute{
  Jefferson Lab,
  12000 Jefferson Ave,
  Newport News VA 23606,
  USA
  \texttt{flemingg@jlab.org}
}

\maketitle

\section{\label{sec:lqcd_nmr_comparison}Lattice QCD and NMR Spectroscopy}

The Lattice QCD (LQCD) community has occasionally gone through periods
of self-examination of its data analysis methods and compared them
with methods used in other disciplines
\citep{Toussaint:1990,Michael:1994yj,Michael:1995sz}.
This process has shown that the techniques widely used elsewhere
may also be useful in analyzing LQCD data.  It seems that we are
in such a period now with many groups trying what are generally called
Bayesian methods such as Maximal Entropy (MEM) or constrained fitting
\citep[and many others]{Nakahara:1999vy,Lepage:2001ym,Allton:2002mr,
Fiebig:2002sp,Draper:2003cg}.  In these proceedings we will attempt to apply
this process to a comparison of data modeling techniques used in LQCD
and NMR Spectroscopy to see if there are methods which may also be useful
when applied to LQCD data.

A common problem in Lattice QCD is the estimation of hadronic energies
$E_k(\vec{p})$ of $k = 1 \cdots K$ states from samples of the hadronic
correlation function of a specified set of quantum numbers computed
in a Monte Carlo simulation.  A typical model function is
\begin{equation}
  \label{eq:hadron_correlation}
  C(\vec{p}, t_n) = \sum_{k=1}^K A_k(\vec{p})
  \exp\left[-(t_0 + n a) E_k(\vec{p})\right]
\end{equation} \[
  A_k, E_k \in \mathbb{R}, \quad
  0 \le E_1 \le \cdots \le E_k \le E_{k+1} \le \cdots \le E_K
\]
where one of the quantum numbers, the spatial momentum $\vec{p}$, 
is shown explicitly for illustration.  The correlation function is estimated
at each time $t_n$, $n = 0 \cdots N-1$ with the $N$ chosen such that
$( E_2 - E_1 ) t_{N-1} \gg 1$.  This enables the ground state energy
$E_1$ to be easily determined from the large time behavior.  To accurately
estimate the $k$-th energy level requires choosing a sampling interval
$a^{-1} \gg E_k - E_1$.  Unfortunately, computational constraints typically
force us to choose time intervals larger $(a^{-1} \sim 2 \mathrm{\ GeV})$
and number of time samples smaller $(N \sim 32)$ than is ideally preferred.

In an idealized nuclear magnetic resonance (NMR) spectroscopy\footnote{
  In medical applications, MRS is a preferred abbreviation, probably
  to avoid the perceived public aversion to anything \textit{nuclear}.
}
experiment, a sample is placed in an external magnetic field and a transient
field from an RF coil is used to temporarily drive the various nuclei into a
non-equilibrium distribution of magnetic spin states.  Then, as the sample
relaxes back to its equilibrium distribution, each type of excited nuclei
radiates at a characteristic frequency $f_k$.  The sum of all these
microscopic signals are picked up by another RF coil, giving rise to the free
induction decay (FID) signal
\begin{equation}
  \label{eq:FID}
  y_n = \sum_{k=1}^K a_k e^{i \phi_k} e^{(-d_k + i 2 \pi f_k) t_n} + e_n,
  \quad n \in [0,N-1]
\end{equation} \[
  a_k,\ \phi_k,\ d_k,\ f_k \in \mathbb{R}; \quad d_k \ge 0;
  \quad \mathrm{noise:}\ e_n \in \mathbb{C}.
\]
As the frequencies are known \textit{a priori}, an experienced operator can
incorporate this prior knowledge by Fourier transforming the data and matching
Lorentzian peaks against existing databases.  Bayesian methods are then used
to constrain the frequencies, enabling the estimation of the other parameters.
Of particular interest are the amplitudes $a_k$, which are related to the
number of various nuclei in the sample, and the damping rates $d_k$, which are
related to the mobility and molecular environment of the nuclei.

Both Eqs.~(\ref{eq:hadron_correlation}) and (\ref{eq:FID})
can be written in the form
\begin{equation}
\label{eq:vandermonde_system}
y_n = \sum_{k=1}^{K} a_k \alpha_k^n
\end{equation}
or in matrix notation $\bvec{y} = \bvec{\Phi}(\bvec{\alpha})\ \bvec{a}$.
In numerical analysis, this is known as a Vandermonde system
and $\bvec{\Phi}$ is a Vandermonde matrix.  Note also that all the parameters
$\bvec{\alpha}$ which enter non-linearly in the model only appear
in the Vandermonde matrix and the remaining linear parameters in $\bvec{a}$.
This suggests that if the best fit values of only the non-linear parameters,
$\bvec{\widehat{\alpha}}$, were known \textit{a priori} then the remaining
best fit values of the linear parameters, $\bvec{\widehat{a}}$
could be determined using a linear least squares algorithm.  Hence,
linear and non-linear parameters need not be determined simultaneously
and in Sec.~\ref{sec:VARPRO} we will discuss the best known algorithm
that exploits this feature.

We have found that all of the model functions we use to fit
hadronic correlations in LQCD can be written in the Vandermonde form.
For a less trivial example, here is the model function for mesonic
correlations with periodic (or anti-periodic) temporal boundary conditions
and either Wilson ($\sigma$=1) or staggered ($\sigma$=-1) fermions
\begin{equation}
\label{eq:periodic_mesonic_correlation}
C(\tau_n) = \sum_{k=1}^K \sigma^{k n} A_k e^{-a N E_k / 2} \cosh(a n E_k),
\quad 0 \le E_k \le E_{k+2}.
\end{equation}
In this case, if we choose $\alpha_k = \sigma^k \cosh(a E_k)$
to be the parameters of the Vandermonde matrix $\bvec{\Phi}$
then we can construct the data vector $\bvec{y}$ from the correlation data
\begin{equation}
\label{eq:periodic_mesonic_data_vector}
y_n = \frac{1}{2^{n-1}} \sum_{j=0}^{n-1} \binom{n - 1}{j} C(\tau_{n-2j-1}) .
\end{equation}
where $\binom{n}{j}$ are binomial coefficients.

In NMR spectroscopy and in LQCD, fitting data often requires an experienced
user to interact with the fitting program, \textit{i.e.}\ to provide
initial guesses to the minimizer or to choose what prior knowledge
may be used to constrain the minimization, and this can often be
a time-consuming process if the data are of marginal quality.
In LQCD fitting programs, the \textsl{effective mass} technique
is often used to provide non-interactive initial guesses to the minimizer.
In NMR spectroscopy, more general analogues, called \textsl{black box}
methods, have been developed for situations where an expert user
is unavailable or the rate of data acquisition precludes interaction.
In Sec.~\ref{sec:black_box_methods}, we will look at the generalization
of the effective mass technique, which will lead to a Hankel system
that must be solved.

\section{\label{sec:VARPRO}\texttt{VARPRO}: Variable Projection algorithm}

In Sec.~\ref{sec:lqcd_nmr_comparison}, we considered data whose
model may be written as $\bvec{y} = \bvec{\Phi} \bvec{a}$,
as in Eq.~(\ref{eq:vandermonde_system}), with the data vector
$\bvec{y} \in \mathbb{R}^{N}$ and the linear parameter vector
$\bvec{a} \in \mathbb{R}^{K}$ and $N > 2K$ is necessary for the problem
to be over-determined.  The non-linear parameter vector
$\bvec{\alpha}$ is used to determine the components of the non-linear
parameter matrix $\bvec{\Phi} \in \mathbb{R}^{N \times K}$
of the general form
\begin{equation}
\label{eq:Phi_general}
\bvec{\Phi} = \left( \begin{array}{ccc}
\phi_1(t_1, \bvec{\alpha}) & \cdots & \phi_K(t_1, \bvec{\alpha}) \\
\vdots                       & \ddots & \vdots                       \\
\phi_1(t_N, \bvec{\alpha}) & \cdots & \phi_K(t_N, \bvec{\alpha}) \\
\end{array}\right).
\end{equation}
Non-linear least squares problems of this type form a special class
known as separable non-linear least squares and have been well studied
in the numerical analysis community for the past thirty years.

To see how this special structure can be exploited, recall the least squares
functional to be minimized is
\begin{equation}
\label{eq:NLLS_cost_functional}
r_1^2(\bvec{\alpha}, \bvec{a}) = \left|
  \bvec{y} - \bvec{\Phi}(\bvec{\alpha}) \bvec{a}
\right|^2 .
\end{equation}
Now, suppose we were given \textit{a priori} the value of the non-linear
parameters $\bvec{\alpha}$ at the minimum
of Eq.~(\ref{eq:NLLS_cost_functional}) which we denote
$\bvec{\hat{\alpha}}$.  We can easily determine \textit{a posteriori}
the linear parameters $\bvec{\hat{a}}$ by solving the corresponding
\textsl{linear} least squares problem.  The solution is simply
\begin{equation}
\label{eq:LS_solution}
\bvec{\hat{a}} = \bvec{\Phi}^+(\bvec{\hat{\alpha}}) \bvec{y}
\end{equation}
where $\bvec{\Phi}^+(\bvec{\hat{\alpha}})$ is the Moore--Penrose
pseudo-inverse of $\bvec{\Phi}(\bvec{\hat{\alpha}})$ \citep{Weisstein:2004}.
Substituting Eq.~(\ref{eq:LS_solution}) back
into Eq.~(\ref{eq:NLLS_cost_functional}) we get a new least squares
functional that depends only on $\bvec{\alpha}$
\begin{equation}
\label{eq:VARPRO_functional_1}
r_2^2(\bvec{\alpha}) = \left|
  \bvec{y} - \bvec{\Phi}(\bvec{\alpha}) \bvec{\Phi}^+(\bvec{\alpha}) \bvec{y}
\right|^2 .
\end{equation}
$\bvec{P}(\bvec{\alpha}) \equiv \bvec{\Phi}(\bvec{\alpha})
\bvec{\Phi}^+(\bvec{\alpha})$ is the orthogonal projector onto the linear
space spanned by the column vectors of $\bvec{\Phi}(\bvec{\alpha})$, so
$\bvec{P}^\perp(\bvec{\alpha}) \equiv \bvec{1} - \bvec{P}(\bvec{\alpha})$
is the projector onto the orthogonal complement of the column space
of $\bvec{\Phi}(\bvec{\alpha})$.  Hence, we can rewrite
Eq.~(\ref{eq:VARPRO_functional_1}) more compactly as
\begin{equation}
\label{eq:VARPRO_funcitonal_2}
r_2^2(\bvec{\alpha})
= \left| \bvec{P}^\perp(\bvec{\alpha}) \bvec{y} \right|^2 .
\end{equation}
This form makes it easier to see why $r_2^2(\bvec{\alpha})$ is 
commonly called the \textsl{variable projection} (VARPRO) functional.
It has been shown \citep{Golub:1973b} that the minima of $r_2(\bvec{\alpha})$
and the corresponding values of $\bvec{a}$ from Eq.~(\ref{eq:LS_solution})
are equivalent to the minima of $r_1^2(\bvec{\alpha},\bvec{a})$.

One complication of the VARPRO method is computing the gradient
$\partial \bvec{r}_2 / \partial \bvec{\alpha}$ when the gradients
$\partial \phi_k(t_n, \bvec{\alpha}) / \partial \bvec{\alpha}$
are known.  The solution is presented in some detail in \citep{Golub:1973b}
and an excellent \texttt{FORTRAN} implementation \citep{Bolstad:1977}
is available in the Netlib Repository.

From our review of the NMR spectroscopy literature, it appears that
the VARPRO method, and in particular the Netlib implementation,
is competitive with the standard least squares method using either
the \texttt{LMDER} routine of the \texttt{MINPACK} library
or the \texttt{NL2SOL} routines of the \texttt{PORT} library,
both also available in the Netlib Repository.  In general, the VARPRO
functional requires fewer minimizer iterations, but the gradient
computation is more expensive.  Note that the Levenberg-Marquardt minimizer
in \citep{Press:1992} performs quite poorly relative to these three
and we cannot recommend its use in production code.

Apart from the issue of numerical speed and accuracy of the VARPRO method,
we see two additional benefits of this method over the standard method.
First, by reducing the dimensionality of the search space by postponing
the determination of $\bvec{\hat{a}}$, this also means that starting estimates
for $\bvec{\hat{a}}$ are not needed.  For LQCD, this is a great benefit,
since good guesses for $\bvec{\alpha}$ are easily obtained
from the black box methods of Sec.~\ref{sec:black_box_methods}.  Second,
when the incorporation of Bayesian prior knowledge is desired,
for LQCD it seems easier to develop reasonable priors for the energies $E_k$
than the amplitudes $A_k$.  When using the VARPRO method, only priors
for the energies are needed.  Of course, if reliable priors for the amplitudes
are available, one should instead use the standard method.  Finally,
data covariance can easily be incorporated in the usual way
\begin{equation}
r_2^2(\bvec{\alpha}) = \left[ \bvec{P}^\perp(\bvec{\alpha}) \bvec{y} \right]^T
\bvec{C}^{-1}(\bvec{y}) \left[ \bvec{P}^\perp(\bvec{\alpha}) \bvec{y} \right].
\end{equation}

\section{\label{sec:black_box_methods}Black Box Methods}

\subsection{\label{sub:effective_mass}Effective Masses}

The best example of a black box method widely used in LQCD
is the method of effective masses.  Let's consider the problem
of Eq.~(\ref{eq:vandermonde_system}) for the case $N$=2, $K$=1
\begin{equation}
\left( \begin{array}{c} y_n \\ y_{n+1} \end{array} \right) =
\left( \begin{array}{c} \alpha_1^n \\ \alpha_1^{n+1} \end{array} \right)
\left( a_1 \right) \quad \Rightarrow \quad
\alpha_1 = \frac{y_{n+1}}{y_n},
\quad a_1 = \frac{y_n}{\alpha_1^n}
\end{equation}
As expected, the problem is exactly determined, so there is an unique
zero residual solution.  For the model function of Eq.~(\ref{eq:hadron_correlation}) the effective mass is $m_\mathrm{eff} = - \log(\alpha_1)$.  Note that
the non-linear parameter $\alpha_1$ is determined first from the data
and then the linear parameter $a_1$ can be determined.  This is an indication
of the separability of the least squares problem discussed
in Sec.~\ref{sec:VARPRO}.

As we are unaware of its presentation elsewhere, here is the two-state
effective mass solution.  We start from Eq.~(\ref{eq:vandermonde_system})
for $N$=4, $K$=2
\begin{equation}
\label{eq:4x2_vandermonde_system}
\left( \begin{array}{l}
  y_{n} \\ y_{n+1} \\ y_{n+2} \\ y_{n+3}
\end{array} \right) = \left( \begin{array}{ll}
  1          & 1          \\
  \alpha_1   & \alpha_2   \\
  \alpha_1^2 & \alpha_2^2 \\
  \alpha_1^3 & \alpha_2^3
\end{array} \right)
\left( \begin{array}{c} a_1 \alpha_1^n \\ a_2 \alpha_2^n \end{array} \right).
\end{equation}
If we compute three quantities from the data
\begin{eqnarray}
A & = & y_{n+1}^2       - y_{n  } y_{n+2} \\
B & = & y_{n  } y_{n+3} - y_{n+1} y_{n+2} \\
C & = & y_{n+2}^2       - y_{n+1} y_{n+3}
\end{eqnarray}
then the two solutions for the non-linear parameters $\alpha_k$
come from the familiar quadratic equation
\begin{equation}
\label{eq:n=4_k=2_nonlin_solution}
\alpha_{1,2} = \frac{-B \pm \sqrt{B^2-4AC}}{2A} .
\end{equation}
As before, the linear parameters $a_{1,2}$ can also be determined
once the non-linear parameters are known
\begin{equation}
\label{eq:n=4_k=2_lin_solution}
a_k \alpha_k^n = \frac{1}{2} \left[
  y_n \pm \frac{\sqrt{(B^2-4AC)[4A^3+(B^2-4AC)y_n^2]}}{B^2-4AC}
\right]
\end{equation}
where some care must be taken to properly match solutions.

In general, when $N$=2$K$ there should always be such a unique zero residual
solution.  From inspection of Eq.~(\ref{eq:4x2_vandermonde_system})
the $N$=4, $K$=2 problem is a set of 4 coupled cubic equations.
Unfortunately, due to Abel's Impossibility Theorem \citep{Abel:1826},
we should expect that general algebraic solutions are only possible
for $N$$\le$5.  Yet, the rather surprising result
of Eq.(\ref{eq:n=4_k=2_nonlin_solution}) is that after properly separating
the non-linear parameters $\alpha_k$, the $N$=4, $K$=2 problem is
of quadratic order.  Thus, we suspect that it is also possible to find
algebraic solutions to the three-state and four-state effective mass
problems when properly reduced to cubic and quartic equations
after separation of variables.

\subsection{\label{sub:linear_prediction}Black Box I: Linear Prediction}

In order to compute solutions of Eq.~(\ref{eq:vandermonde_system})
when the system is over-determined ($N>2K$) or when an algebraic
solution is not available, we consider the first black box method
called \textsl{linear prediction}.  We form a $K$-th order polynomial
with the $\alpha_k$ as roots
\begin{equation}
\label{eq:forward_linear_prediction_coeff}
p(\alpha) = \prod_{k=1}^K (\alpha - \alpha_k)
= \sum_{i=0}^K p_i \alpha^{K-i} \quad (p_0 = 1).
\end{equation}
Since $p(\alpha_k) = 0$ the following is true
\begin{equation}
\label{eq:large_powers_of_roots}
\alpha_k^m = - \sum_{i=1}^K p_i\ \alpha_k^{m-i},\quad m \ge K.
\end{equation}
When Eq.~(\ref{eq:large_powers_of_roots}) is substituted
in Eq.~(\ref{eq:vandermonde_system}) we find the following relation
\begin{equation}
\label{eq:forward_linear_prediction}
y_m = - \sum_{k=1}^K p_k\ y_{m-k}, \quad m \ge K.
\end{equation}
Because Eq.~(\ref{eq:forward_linear_prediction}) enables us to ``predict''
the data $y_m$ at larger times in terms of the data
$y_{m-K}, \cdots, y_{m-1}$ at earlier times, the $p_k$ are commonly called
\textsl{forward linear prediction coefficients}.

Using Eq.~(\ref{eq:forward_linear_prediction}) we can construct
the linear system $\bvec{h}_\mathrm{lp} = -\bvec{H}_\mathrm{lp} \bvec{p}$
\begin{equation}
\label{eq:linear_prediction_hankel_system}
\left( \begin{array}{c}
  y_{M} \\ y_{M+1} \\ \vdots \\ y_{M-1}
\end{array} \right) = - \left( \begin{array}{ccc}
  y_0       & \cdots & y_{M-1} \\
  y_1       & \cdots & y_{M}   \\
  \vdots    & \ddots & \vdots  \\
  y_{N-M-1} & \cdots & y_{N-2}
\end{array} \right) \left( \begin{array}{c}
  p_M \\ p_{M-1} \\ \vdots \\ p_1
\end{array} \right),
\quad N \ge 2M.
\end{equation}
In numerical analysis, this is known as a Hankel system and the matrix
$\bvec{H}_\mathrm{lp}$ is a Hankel matrix.  After solving
Eq.~(\ref{eq:linear_prediction_hankel_system}) for $\bvec{p}$, the roots
of the polynomial of Eq.~(\ref{eq:forward_linear_prediction_coeff})
are computed to determine the parameters $\alpha_k$.  The $a_k$ parameters
can subsequently be determined from Eq.~(\ref{eq:LS_solution}).

In the presence of noisy data, the equality
in Eq.~(\ref{eq:linear_prediction_hankel_system}) is only approximate,
even for the case $N=2M$, so some minimization method like least
squares must be used.  This doesn't mean that the parameter estimates
from linear prediction agree with the parameter estimates from the
least squares methods of Sec.~\ref{sec:VARPRO}.  Gauss proved
\citep{Gauss:1823} that the least squares estimates of fit parameters
for linear problems have the smallest possible variance.  In this sense,
least squares estimates are considered \textsl{optimal} although we know
of no proof that this holds for non-linear problems.  Since linear
prediction estimates may not agree with least squares, they are considered
\textsl{sub-optimal} even though there is no proof that the variance
is larger, due to non-linearity.

A popular method for solving Eq.~(\ref{eq:linear_prediction_hankel_system})
is the LPSVD algorithm \citep{Kumeresan:1982}.  In this method,
we construct $\bvec{H}_\mathrm{lp}$ for $M$ as large as possible,
even if we are only interested in estimating $K < M$ parameters.
After computing the SVD of $\bvec{H}_\mathrm{lp}$, we construct
a rank $K$ approximation $\bvec{H}_{\mathrm{lp}K}$
\begin{equation}
\bvec{H}_\mathrm{lp} = \bvec{U} \bvec{\Sigma} \bvec{V}^\dagger =
\left( \bvec{U}_K \bvec{U}_2 \right) \left( \begin{array}{cc}
  \bvec{\Sigma}_K &                 \\
                  & \bvec{\Sigma}_2 \\
\end{array} \right) \left( \bvec{V}_K \bvec{V}_2 \right)^\dagger ,
\quad
\bvec{H}_{\mathrm{lp}K} = \bvec{U}_K \bvec{\Sigma}_K \bvec{V}_K^\dagger
\end{equation}
$\bvec{\Sigma}_K$ contains the $K$ largest singular values.
By zeroing $\bvec{\Sigma}_2$ to reduce the rank of $\bvec{H}_\mathrm{lp}$,
much of the statistical noise is eliminated from the problem.
From the Eckart--Young--Mirsky theorem \citep{Eckart:1936,Mirsky:1960},
this rank $K$ approximation is the nearest possible under either
the Frobenius norm or matrix 2-norm.
Then, after solving $\bvec{h}_\mathrm{lp} = -\bvec{H}_{\mathrm{lp}K} \bvec{p}$
for the $p_m$ coefficients, the $M$ roots of the polynomial
in Eq.~(\ref{eq:forward_linear_prediction_coeff}) are computed
using a root-finding algorithm.  Since the rank of $\bvec{H}_\mathrm{lp}$
was reduced to $K$, only $K$ roots are valid parameter estimates.
Typically, the $K$ largest magnitude roots are chosen.

Our experience with this algorithm is that the largest magnitude
roots often have unphysical values if $K$ is set larger
than a reasonable number given the statistical precision of the data.
There are also several issues which may be of some concern.
First, we found that root-finding algorithms
all come with caveats about stability and susceptibility
to round-off error and must be treated with some care.
Also, since statistical noise is present on both sides
of Eq.~(\ref{eq:linear_prediction_hankel_system}), the rank-reduced
least squares solution is probably not appropriate and one should
probably use an errors-in-variables (EIV) approach
like total least squares (TLS), which we will describe
in Sec.~\ref{sub:total_least_squares}.  We have found that the TLS variant
of LPSVD, called LPTLS \citep{Tirendi:1989}, gives better parameter estimates
than LPSVD.

\subsection{\label{sub:total_least_squares}Total Least Squares}

In the standard linear least squares problem
$\bvec{A}\bvec{x} \approx \bvec{b}$
\begin{equation}
\label{eq:LS_problem}
\underset{\bvec{x} \in \mathbb{R}^K}{\mathrm{minimize}}
\left\| \bvec{A}\bvec{x} - \bvec{b} \right\|_2,
\qquad
\bvec{A} \in \mathbb{R}^{N \times K},
\ \bvec{b} \in \mathbb{R}^N,
\ N \ge K
\end{equation}
an important assumption for the solution,
\textit{i.e.}\ Eq.~(\ref{eq:LS_solution}), to be considered optimal
is that the only errors are in the data vector $\bvec{b}$
and further that those errors are \textsl{i.i.d.}\ (independent and
identically distributed).  When errors also occur in $\bvec{A}$,
as in Eq.~(\ref{eq:linear_prediction_hankel_system}), then a new approach,
often called errors-in-variables (EIV), is required to restore optimality.
Note that the errors in $\bvec{A}$ that cause the loss of optimality
need not be purely statistical: numerical round-off errors
or choosing to fit a model function which differs from the ``true''
model function are potential sources of error which could cause
loss of optimality.

To understand the total least squares (TLS) solution to the EIV
problem, consider the case when a zero residual solution
to Eq.~(\ref{eq:LS_problem}) exists.  Then, if we add $\bvec{b}$ as
a column of $\bvec{A}$, written $\left[ \bvec{A} \bvec{b} \right]$,
it cannot have greater column rank than $\bvec{A}$ because
$\bvec{b} \in \mathrm{Ran}(\bvec{A})$.  If we compute the SVD
of $\left[ \bvec{A} \bvec{b} \right]$ we will find
that the singular value $\sigma_{K+1} = 0$.  When the solution
of Eq.~(\ref{eq:LS_problem}) has non-zero residual, we may find
the singular value $\sigma{K+1}$ of $\left[ \bvec{A} \bvec{b} \right]$
to be non-zero as well. But, we can construct the nearest rank $R \le K$
approximation to $\left[ \bvec{A} \bvec{b} \right]$ (in the sense
of the Eckart--Young--Mirsky theorem) and this gives us the TLS solution.
The TLS solution was computed in \citep{Golub:1970,Golub:1973a},
although the name was not coined until \citep{Golub:1980}.
A comprehensive review \citep{VanHuffel:1991b} of the subject is available.

Finally, TLS is very sensitive to the distribution of errors
in $\left[\bvec{A}\bvec{b}\right]$.  If the errors are not known
to be i.i.d.\ then it is crucial to scale the matrix before using
the TLS algorithm.  If the data are uncorrelated, then a method
known as ``\textsl{equilibrium}'' scaling \citep{Bauer:1963}
is sufficient.  If the data are correlated, then Cholesky factors
of the covariance matrix must be used.  In this case, it is better
to use either the generalized TLS algorithm (GTLS)
\citep{VanHuffel:1990a,VanHuffel:1990b}
or the restricted TLS algorithm (RTLS) \citep{VanHuffel:1991a}
which are more robust when the covariance matrix is ill-conditioned.
Implementations of various TLS algorithms are available
in the Netlib Repository \citep{VanHuffel:1988}.

\subsection{\label{sub:state_space}Black Box II: State Space Methods}

The name for these methods is derived from state-space theory
in the control and identification literature \citep{Kung:1983}.
The basic approach is to compute the non-linear parameters $\alpha_k$
of Eq.~(\ref{eq:vandermonde_system}) without needing to compute
the roots of a polynomial, as in Sec.~\ref{sub:linear_prediction}.
From Eq.~(\ref{eq:linear_prediction_hankel_system}), we start by noting
that $\bvec{H}_s = \left[ \bvec{H}_\mathrm{lp} \bvec{h}_\mathrm{lp} \right]$
is also a Hankel matrix
\begin{equation}
\label{eq:state_space_hankel_matrix}
\bvec{H}_s = \left( \begin{array}{ccc|c}
  y_{0    } & \cdots & y_{M-1} & y_{M  } \\
  \vdots    & \ddots & \vdots  & \vdots  \\
  y_{N-M-1} & \cdots & y_{N-2} & y_{N-1} \\
\end{array} \right)
\qquad
M \ge K,\ N-M > K
\end{equation}
A Vandermonde decomposition exists for this matrix
\begin{equation}
\label{eq:state_space_vandermonde_decomposition}
\bvec{S}\bvec{A}\bvec{T}^T = \left( \begin{array}{ccc}
  1                & \cdots & 1                \\
  \alpha_1         & \cdots & \alpha_K         \\
  \vdots           & \ddots & \vdots           \\
  \alpha_1^{N-M-1} & \cdots & \alpha_K^{N-M-1}
\end{array} \right) \left( \begin{array}{ccc}
  a_1 &        &     \\
      & \ddots &     \\
      &        & a_K
\end{array} \right) \left( \begin{array}{ccc}
  1                & \cdots & 1                \\
  \alpha_1         & \cdots & \alpha_K         \\
  \vdots           & \ddots & \vdots           \\
  \alpha_1^{M    } & \cdots & \alpha_K^{M    }
\end{array} \right)^T
\end{equation}
in terms of the linear ($a_k$) and non-linear ($\alpha_k$) parameters
of Eq.~(\ref{eq:vandermonde_system}).  If we could compute this decomposition
directly, then the problem would be solved.  Alas, no such algorithm
is currently known.

An indirect method exists to compute this decomposition called
Hankel SVD (HSVD).  We will consider here a TLS variant called
HTLS \citep{VanHuffel:1994}.  First, we note the shift invariance
property of $\bvec{S}$ (and similarly for $\bvec{T}$)
\begin{equation}
\bvec{S}^\uparrow \bvec{\mathcal{A}} = \bvec{S}_\downarrow,
\qquad
\bvec{\mathcal{A}} = \bvec{\mathrm{diag}}(\alpha_1, \cdots, \alpha_K).
\end{equation}
Next, we note that if such a decomposition is possible,
then $\bvec{S}$, $\bvec{A}$ and $\bvec{T}$ are all of rank $K$
by inspection, so $\bvec{H}_s$ is at least of rank $K$, as well.
So, using SVD we construct the nearest rank $K$ approximation
to $\bvec{H}_{sK}$
\begin{equation}
\label{eq:state_space_hankel_matrix_svd}
\bvec{H}_{sK} = \left( \bvec{U}_K \bvec{U}_2 \right) \left( \begin{array}{cc}
  \bvec{\Sigma}_K & \\ & 0
\end{array} \right) \left( \bvec{V}_K \bvec{V}_2 \right)^\dagger
= \bvec{U}_K \bvec{\Sigma}_K \bvec{V}_K^\dagger
\end{equation}
By comparing the decompositions
of Eq.~(\ref{eq:state_space_vandermonde_decomposition})
and Eq.~(\ref{eq:state_space_hankel_matrix_svd}) we can see
\begin{equation}
\mathrm{Span}(\bvec{S}) = \mathrm{Span}(\bvec{U}_K)
\ \implies\ \bvec{U}_K = \bvec{S} \bvec{Q}
\ \implies\ \bvec{U}_K^\uparrow = \bvec{U}_{K\downarrow}
    \bvec{Q}^{-1} \bvec{\mathcal{A}} \bvec{Q}
\end{equation}
So, computing the TLS solution
of $\left[ \bvec{U}_K^\uparrow \bvec{U}_{K\downarrow} \right]$
will give us $\bvec{Q}^{-1} \bvec{\mathcal{A}} \bvec{Q}$,
which we can then diagonalize using an eigenvalue solver
to get our estimates of $\alpha_k$.

In our experience with these black box methods, the HTLS algorithm
seems to be the most robust.  However, we would like to emphasize
two points.

First, the estimates of $\alpha_k$ from HTLS are considered
sub-optimal because $\bvec{H}_{sK}$
in Eq.~(\ref{eq:state_space_hankel_matrix_svd}) is only approximately,
but not \textsl{exactly}, a Hankel matrix because the SVD
does not enforce the Hankel structure throughout.  A similar
problem occurs while constructing the TLS solution
of $\left[ \bvec{U}_K^\uparrow \bvec{U}_{K\downarrow} \right]$.
\textsl{Structured} TLS algorithms (STLS) exist which can construct
$\bvec{H}_{sK}$ while preserving the Hankel structure
(see \citep{Vanhamme:1999} for references) and hence restoring
the optimality of the estimates.  While we have not yet tried these
STLS algorithms, we note that all of them involve iterative
procedures to restore the structure.
Thus, under the \textsl{``no free lunch''}
theorem, we suspect that the price of restoring optimality
is roughly equivalent to performing the (optimal) non-linear least squares
minimizations described in Sec.~\ref{sec:VARPRO}.

Our second observation is that LQCD data is always correlated,
so that a GTLS or RTLS algorithm is needed to compute the TLS solution
of $\left[ \bvec{U}_K^\uparrow \bvec{U}_{K\downarrow} \right]$.
But, covariance estimates of $\bvec{U}_K$ are not readily computed
from the data covariance matrix because of the required SVD.  Thus,
a jackknife or bootstrap resampling method is required to estimate
$\bvec{\mathrm{cov}}(\bvec{U}_K)$.  Since we expect to use
a resampling method to estimate the covariance of the $\alpha_k$,
this means that there is an inner and outer resampling loop
so the method can easily become computationally expensive
if the number of data samples becomes large.  In this case,
blocking the data is recommended.

\section{\label{sec:conclusions}Conclusions}

We have found that reviewing the literature of other fields
where data analysis of exponentially damped time series
is also prevalent to be quite fruitful.  Our review has discovered
several mature analysis methods which are virtually unknown
(or unmentioned) in the Lattice QCD literature.  We have performed
several tests of all the methods discussed on fake data and on some actual
LQCD data are encouraged by the results. So, we are incorporating
these techniques into our production versions of analysis programs
and expect to report results soon.

Finally, we would like to acknowledge that we have found
Leentje Vanhamme's Ph.D.~Thesis \citep{Vanhamme:1999} an extremely useful
guide to the literature of the NMR spectroscopy community.
We would encourage anyone interested in learning more to start there.
An electronic copy is currently available online.

This work was supported in part by DOE contract DE-AC05-84ER40150
under which the Southeastern Universities Research Association (SURA)
operates the Thomas Jefferson National Accelerator Facility.

\bibliography{Fleming}

\printindex 

\end{document}